\documentclass[sigconf,screen]{acmart}

\usepackage{listings}
\usepackage[capitalise,nameinlink]{cleveref}
\usepackage[export]{adjustbox}
\usepackage[many]{tcolorbox}
\usepackage{subcaption}
\usepackage{tikz}
\usepackage{enumitem}
\usepackage[linesnumbered,ruled]{algorithm2e}
\usepackage{setspace}

\usetikzlibrary{shapes.geometric, arrows.meta, positioning, fit, calc}

\newcommand{\C}[1]{\texttt{\small#1}}

\crefformat{section}{#2\S{}#1#3}
\crefname{algocf}{Alg.}{Algorithms}

\usetikzlibrary{shapes.geometric, arrows.meta, positioning, fit, calc}

\definecolor{RoyalBlue}{HTML}{0071BC}
\definecolor{ForestGreen}{HTML}{009B55}
\definecolor{LinkColor}{rgb}{0.55,0.0,0.3}
\definecolor{CiteColor}{rgb}{0.55,0.0,0.3}
\definecolor{HighlightColor}{rgb}{0.0,0.0,0.0}
\definecolor{Gray}{gray}{0.9}
\definecolor{grey}{rgb}{0.5,0.5,0.5}
\definecolor{red}{rgb}{1,0,0}
\definecolor{darkgreen}{rgb}{0.0,0.7,0.0}
\definecolor{backcolor}{rgb}{0.95,0.95,0.95}

\tcolorboxenvironment{lstlisting}{
  spartan,
  frame empty,
  boxsep=0mm,
  left=0mm,right=0mm,top=-1mm,bottom=-1mm,
  colback=backcolor,
}
\lstdefinestyle{mystyle}{
  keywordstyle=\color{RoyalBlue},
  commentstyle=\color{ForestGreen},
  basicstyle=\linespread{0.9}\ttfamily\footnotesize,
  breakatwhitespace=false,
  breaklines=true,
  captionpos=b,
  keepspaces=true,
  numbers=none,
  numbersep=1pt,
  numberstyle=\scriptsize,
  showspaces=false,
  showstringspaces=false,
  escapechar=$,
}
\lstdefinelanguage{Rust}{
  keywords={as,break,const,continue,crate,else,enum,extern,false,fn,for,if,
  impl,in,let,loop,match,mod,move,mut,pub,ref,return,self,Self,static,struct,
  super,trait,true,type,unsafe,use,where,while,async,await,dyn,abstract,become,
  box,do,final,macro,override,priv,typeof,unsized,virtual,yield,try,union},
  morecomment=[l]{//},
  morecomment=[s]{/*}{*/},
}

\AtBeginDocument{%
  }

\copyrightyear{2024}
\acmYear{2024}
\setcopyright{rightsretained}
\acmConference[ASE '24]{39th IEEE/ACM International Conference on Automated Software Engineering }{October 27-November 1, 2024}{Sacramento, CA, USA}
\acmBooktitle{39th IEEE/ACM International Conference on Automated Software Engineering (ASE '24), October 27-November 1, 2024, Sacramento, CA, USA}
\acmDOI{10.1145/3691620.3694985}
\acmISBN{979-8-4007-1248-7/24/10}

\begin{document}

\title
[To Tag, or Not to Tag: Translating C's Unions to Rust's Tagged Unions]
{To Tag, or Not to Tag:\\Translating C's Unions to Rust's Tagged Unions}


\author{Jaemin Hong}
\orcid{0000-0003-4067-7369}
\affiliation{%
  \institution{KAIST}
  \city{Daejeon}
  \country{South Korea}
}
\email{jaemin.hong@kaist.ac.kr}

\author{Sukyoung Ryu}
\orcid{0000-0002-0019-9772}
\affiliation{%
  \institution{KAIST}
  \city{Daejeon}
  \country{South Korea}
}
\email{sryu.cs@kaist.ac.kr}



\begin{abstract}
  Automatic C-to-Rust translation is a promising way to enhance the reliability
  of legacy system software.
  However, C2Rust, an industrially developed translator, generates Rust code
  with unsafe features, undermining the translation's objective.
  While researchers have proposed techniques to remove unsafe features in
  C2Rust-generated code, these efforts have targeted only a limited subset of
  unsafe features.
  One important unsafe feature remaining unaddressed is a \emph{union}, a type
  consisting of multiple fields sharing the same memory storage.
  Programmers often place a union with a \emph{tag} in a struct to record the
  last-written field, but they can still access wrong fields.
  In contrast, Rust's \emph{tagged unions} combine tags and unions at the
  language level, ensuring correct value access.
  In this work, we propose techniques to replace unions with tagged unions
  during C-to-Rust translation.
  We develop a static analysis that facilitates such replacement by identifying
  tag fields and the corresponding tag values.
  The analysis involves a must-points-to analysis computing struct field values
  and a heuristic interpreting these results.
  To enhance efficiency, we adopt intraprocedural function-wise analysis,
  allowing selective analysis of functions.
  Our evaluation on 36 real-world C programs shows that the proposed approach is
  (1) precise, identifying 74 tag fields with no false positives and only five
  false negatives,
  (2) mostly correct, with 17 out of 23 programs passing tests
  post-transformation, and
  (3) efficient, capable of analyzing and transforming 141k LOC in 4,910 seconds.
\end{abstract}

\ccsdesc[500]{Software and its engineering~Source code generation}
\ccsdesc[300]{Software and its engineering~Automated static analysis}
\ccsdesc[100]{Software and its engineering~Maintaining software}
\ccsdesc[100]{Software and its engineering~Software evolution}

\keywords{Rust, C, Automatic Translation, Union, Tagged Union}


\maketitle

\section{Introduction}
\label{sec:intro}

Translating C code to Rust is a promising approach to enhancing the reliability
of legacy system software.
C Programs often suffer from memory bugs leading to critical security
vulnerabilities due to the absence of language-level mechanisms to prevent
them~\cite{chen2011linux, msrcblog}.
Rust, a recently developed system programming language, ensures memory safety
at compile time through type checking~\cite{matsakis2014rust, jung2017rustbelt}.
By translating legacy C code to Rust, developers can detect previously unknown
bugs and prevent introducing new bugs~\cite{rust-curl}.

Since manual translation is laborious and error-prone, an automatic C-to-Rust
translator named C2Rust~\cite{c2rust} has been developed in the industry.
It converts C code to Rust by leveraging \emph{Unsafe Rust}~\cite{unsafe-rust},
which allows the use of \emph{unsafe} language features.
These features, such as dereferencing raw pointers and calling functions in
external code, are equivalent to C's features and enable straightforward
syntactic translation.
However, as the compiler does not ensure their safety, their use contradicts the
goal of translation.

To address this, researchers have proposed techniques to reduce the use of
unsafe features in C2Rust-generated code by replacing them with safe
counterparts in Rust.
\textsc{Laertes}~\cite{emre2021translating, emre2023aliasing} and
\textsc{Crown}~\cite{zhang2023ownership} replace raw pointers with references,
whose validity is guaranteed by the compiler.
Concrat~\cite{hong2023concrat} replaces certain external function calls by
substituting the C lock API with the Rust lock API.
Unfortunately, raw pointers and external functions are not the only sources of
unsafety, and previous studies have neglected other unsafe features, limiting
applicability to the translation of real-world C code.

\emph{Unions} are an important source of unsafety in C-to-Rust translation that
has not been studied yet.
A union is a compound data type consisting of multiple fields sharing the same
memory storage, facilitating efficient memory use by allowing values of
different types to be stored at the same location~\cite{rust-union}.
Since memory efficiency is crucial in system software, unions are widely used in
C.
Notably, Emre et al.~\cite{emre2021translating} show that 18\% of unsafe
functions (functions using unsafe features) in C2Rust-generated code involve
unions.

Reading a union field is an unsafe feature in Rust because unions do not record
which field has been written to.
If a program reads a field other than the last-written one, the value is
\emph{reinterpreted} as another type.
While reinterpretation is useful for some uses, like packet parsing, it is
dangerous in general.
For example, reinterpreting an integer as a pointer can lead to invalid memory
access.
Thus, many C programs avoid reinterpretation when using unions.

To use unions without reinterpretation, it is essential to decide which field to
read.
Some programs rely on global context to determine the field, but many use
\emph{tags}, i.e., integer values signifying the last-written fields.
When using tags, a union and a tag are placed in a single struct, and the
program checks the tag before accessing the union's field.
However, tags cannot guarantee the absence of reinterpretation.
Programs may read wrong fields after checking tags or set incorrect tag values
when writing to fields.

Rust directly supports this pattern of combining tags and unions as a language
feature called \emph{tagged unions} (or \emph{enums})~\cite{rust-enum}.
This allows defining a tagged union as a single type by enumerating tags and the
type of a value associated with each tag.
By using tagged unions, programmers can avoid mistakenly reinterpreting values.
To access a value in a tagged union, programs must use \emph{pattern matching},
which checks the tag and provides access to the associated value.
When constructing a tagged union, the compiler ensures that the tag and the
value's type match the type definition.
Thus, tagged unions are a safe feature in Rust, making it desirable to replace
unions accompanied by tags with tagged unions.

\begin{figure}[t]
\begin{tikzpicture}[
  node distance=0.2cm and 0.4cm,
  auto,
  data/.style={
    rectangle,
    align=center,
    minimum height=2em
  },
  comp/.style={
    rectangle,
    draw,
    align=center,
    minimum height=1em
  },
  line/.style={draw, -Latex}
]
  \node [data] (ccode) {\footnotesize C Code};
  \node [comp, right=of ccode] (c2rust) {\footnotesize C2Rust};
  \node [data, right=of c2rust] (rustcode1) {\footnotesize Rust code\\[-0.4em]\footnotesize (unions)};
  \node [comp, right=of rustcode1] (transformer) {\footnotesize Transformer (\cref{sec:transformation})};
  \node [data, right=of transformer] (rustcode2)
    {\footnotesize Rust code\\[-0.4em]\footnotesize (tagged unions)};
  \node [comp, below=of rustcode1] (analyzer) {\footnotesize Analyzer (\cref{sec:analysis})};
  \node [data, below=of transformer] (result) {\footnotesize Analysis result};
  \path [line] (ccode) -- (c2rust);
  \path [line] (c2rust) -- (rustcode1);
  \path [line] (rustcode1) -- (transformer);
  \path [line] (transformer) -- (rustcode2);
  \path [line] ($(rustcode1.south)!0.1!(rustcode1.north)$) -- (analyzer);
  \path [line] (analyzer) -- (result);
  \path [line] ($(result.south)!0.8!(result.north)$) -- (transformer);
  \node[draw, inner xsep=6pt, inner ysep=1pt, fit=(c2rust) (rustcode1) (transformer) (analyzer) (result)] (largebox) {};
\end{tikzpicture}
  \vspace{-2em}
\caption{The workflow of the proposed approach}
  \vspace{-1em}
\label{fig:workflow}
\end{figure}
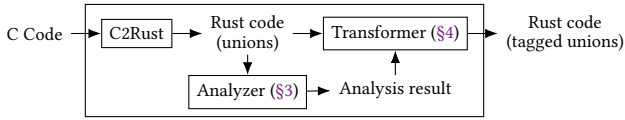

In this work, we propose techniques to translate C's unions to Rust's tagged
unions.
\cref{fig:workflow} shows the workflow of the proposed approach.
We first translate C code to Rust code that still contains unions using C2Rust.
We then transform the C2Rust-generated code by replacing unions with tagged
unions.
To enable this transformation, we perform static analysis to obtain information
related to unions:
(1) the \emph{tag field} (the field containing a tag value) for each union and
(2) tag values associated with each union field.
This static analysis must meet several challenging requirements.

First, the analysis needs to determine the values of struct fields.
Programs typically use \C{switch}/\C{if} to access different union fields in
different branches, using the tag field in the condition.
If a struct field has distinct values when accessing different union fields, it
is likely a tag field, and each distinct value is associated with the accessed
union field.
If it has the same value when accessing different union fields, it is not a tag
field.
Thus, we can identify tag fields by deciding the value of each struct field in
\C{switch}/\C{if} branches.

To achieve this goal, we propose a \emph{must-points-to analysis} capable of
tracking \emph{integer equality}.
To determine a struct's field value at each program point based on the branch
that the program point belongs to, the struct should be the same as the struct
whose field is used in the \C{switch}/\C{if} condition.
Since structs are often passed as pointers, deciding whether they are the same
requires must-points-to relations.
Additionally, programs sometimes use a local variable storing a field's value in
a condition.
In such cases, determining the field's value in each branch requires the
knowledge that the field and the local variable have the same integer.

The second requirement for the analysis is efficiency.
The key idea for achieving efficiency is to selectively analyze functions.
To identify tag fields, we need the field values only of the structs containing
unions.
It means that functions not accessing such structs do not need to be analyzed.
Therefore, we adopt intraprocedural function-wise analysis, instead of
interprocedural whole-program analysis, allowing only the selected functions to
be analyzed.

The third requirement is the ability to identify tag values associated with each
union field despite the imprecision of the analysis.
Given that imprecision is a fundamental limitation of static
analysis~\cite{rice1953classes}, it may not be possible to determine the field
values at some program points.
To address this, we propose a heuristic to interpret such partial information.
The heuristic involves two steps.
First, we examine the accessed union fields and the struct field values in
\C{switch}/\C{if} branches.
This provides reliable information because programs typically access the correct
union field after checking the tag.
However, due to imprecision, it may fail to identify some tag values.
Second, we inspect the last-written union fields and the struct field values at
each program point.
This can capture information missed by the first step but may be incorrect
because of an \emph{intermediate} state, where only the tag or the union field
has been set.
Therefore, we ignore the field associated with a tag by the second step
if it differs from the one associated with the tag by the first step.

Overall, the contributions of this work are as follows:
\vspace{-0.5em}
\begin{itemize}
  \item
    We propose static analysis that identifies tag fields and tag values
    associated with union fields, consisting of must-points-to analysis capable
    of tracking integer equality and a heuristic interpreting the analysis
    results (\cref{sec:analysis}).
  \item
    We propose code transformation replacing unions with tagged unions using the
    analysis results (\cref{sec:transformation}).
  \item
    We implement the proposed approach in a tool named Urcrat ({\bf u}nion-{\bf
    r}emoving {\bf C}-to-{\bf R}ust {\bf a}utomatic {\bf t}ranslator) and
    evaluate it using 36 real-world C programs.
    Our evaluation shows that the approach is
    (1) precise, identifying 74 tag fields with no false positives and only five
    false negatives,
    (2) mostly correct, with 17 out of 23 programs passing tests after
    transformation, and
    (3) efficient, capable of analyzing and transforming 141k LOC in 4,910
    seconds (\cref{sec:evaluation}).
\end{itemize}
\vspace{-0.5em}
We also discuss related work (\cref{sec:related}) and conclude the paper
(\cref{sec:conclusion}).

\section{Background}
\label{sec:background}

In this section, we briefly describe the use of unions with tags in C
(\cref{sec:background:union}), how C2Rust translates such C code to Rust
(\cref{sec:background:c2rust}), and how tagged unions in Rust can safely
represent the same logic (\cref{sec:background:tagged-union}).

\subsection{Unions with Tags}
\label{sec:background:union}

As an example of C code using unions, we use the syntax and evaluation of simple
arithmetic expressions defined as follows:
\begin{center}
  $ e\ ::=\ 1\ \ |\ \ -e\ \ |\ \ e+e\ \ |\ \ e*e $
\end{center}
An expression is either a constant $1$, a negation, an addition, or a
multiplication. This syntax is implemented as follows:

\vspace{-0.5em}
\begin{lstlisting}[language=C]
struct Expr {
  int kind; union { struct Expr *e; struct BExpr b; } v;
};
struct BExpr { struct Expr *l; struct Expr *r; };
\end{lstlisting}
\vspace{-0.5em}

\noindent
\C{struct Expr} is the type of an expression, and its field \C{kind} indicates
the kind of expression:
\C{0} for constant $1$, \C{1} for negation, \C{2} for addition, and \C{3} for
multiplication.
The inner union value \C{v} stores the necessary data for each kind of
expression.
When \C{kind} is \C{1}, the operand of negation is stored in \C{v.e};
when \C{kind} is \C{2} or \C{3}, the operands are stored in \C{v.b.l} and
\C{v.b.r}.
Since \C{kind} signifies which union field has been written to, it is the tag
field for the union.

A function evaluating an expression is implemented as follows:

\vspace{-0.5em}
\begin{lstlisting}[language=C]
int eval(struct Expr *e) {
  switch (e->kind) {
    case 0: return 1;
    case 1: return -eval(e->v.e);
    case 2: return eval(e->v.b.l) + eval(e->v.b.r);
\end{lstlisting}
\begin{lstlisting}[language=C]
    case 3: return eval(e->v.b.l) * eval(e->v.b.r);
    default: exit(1); }}
\end{lstlisting}
\vspace{-0.5em}

\noindent
It evaluates the expression by checking the \C{kind} field and accessing the
appropriate union field accordingly.
When \C{kind} is \C{1}, it accesses the union field \C{e};
when \C{kind} is \C{2} or \C{3}, it accesses \C{b}.
No field is accessed otherwise.

Programs sometimes use \C{if} to check tags, particularly to compare with a
specific tag value.
An example using \C{if} is shown below:

\vspace{-0.5em}
\begin{lstlisting}[language=C]
if (e->kind == 1) return -eval(e->v.e);
\end{lstlisting}
\vspace{-0.5em}

While tag fields are beneficial for accessing the correct union fields, they
cannot prevent memory bugs.
For example, the following code accesses \C{e->v.b} despite \C{kind} being
\C{1}:

\vspace{-0.5em}
\begin{lstlisting}[language=C]
switch (e->kind) {
  case 1: return eval(e->v.b.l) + eval(e->v.b.r);
\end{lstlisting}
\vspace{-0.5em}

\noindent
Here, \C{e->v.b.l} accesses the pointer stored in \C{e->v.e}, but \C{e->v.b.r}
reads an arbitrary value, potentially causing invalid memory access.
Furthermore, even when \C{eval} is correctly implemented, an incorrect tag value
can be assigned during the construction of an \C{Expr}.
The following code sets \C{e.kind} to \C{2} but writes to \C{e.v.e}:

\vspace{-0.5em}
\begin{lstlisting}[language=C]
struct Expr e; e.kind = 2; e.v.e = ...;
\end{lstlisting}
\vspace{-0.5em}

\noindent
Passing a pointer to \C{e} to \C{eval} results in invalid memory access.

\subsection{C2Rust's Translation}
\label{sec:background:c2rust}

C2Rust translates the definition of \C{struct Expr} to Rust as follows:

\vspace{-0.5em}
\begin{lstlisting}[language=Rust]
struct Expr { kind: i32, v: C2RustUnnamed }
union C2RustUnnamed { e: *mut Expr, b: BExpr }
struct BExpr { l: *mut Expr, r: *mut Expr }
\end{lstlisting}
\vspace{-0.5em}

\noindent
Since Rust does not support anonymous types, C2Rust gives the name
\C{C2RustUnnamed} to the union.
If a file contains multiple anonymous types, they get
\C{C2RustUnnamed\_}$n$ where $n$ is a unique integer.

The function \C{eval} is translated as follows:

\vspace{-0.5em}
\begin{lstlisting}[language=Rust]
fn eval(e: *mut Expr) -> i32 {
  match (*e).kind {
    0 => return 1,
    1 => return -eval((*e).v.e),
    2 => return eval((*e).v.b.l) + eval((*e).v.b.r),
    3 => return eval((*e).v.b.l) * eval((*e).v.b.r),
    _ => exit(1), }}
\end{lstlisting}
\vspace{-0.5em}

\noindent
Since Rust provides \C{match} statements instead of \C{switch}, the function
employs \C{match} to check the tag.
While \C{match} is mainly used for pattern matching on tagged unions, it can
also handle integers, similar to \C{switch}, but without fall-through behavior.
%

\subsection{Tagged Unions}
\label{sec:background:tagged-union}

We can implement the same syntax using tagged unions as follows:

\vspace{-0.5em}
\begin{lstlisting}[language=Rust]
struct Expr { v: C2RustUnnamed }
enum C2RustUnnamed {
  One, Neg(*mut Expr), Add(BExpr), Mul(BExpr) }
struct BExpr { l: *mut Expr, r: *mut Expr }
\end{lstlisting}
\vspace{-0.5em}

\noindent
In Rust, the \C{enum} keyword defines tagged unions.
Although the name \C{C2RustUnnamed} is impractical, we retain it for consistency
with the C code.
A tagged union's definition lists its \emph{variants}, i.e., values with
distinct tags.
Each tag is an identifier, not an integer, and the type of a value associated
with each tag is specified after the tag.
The defined tagged union has four variants with tags \C{One}, \C{Neg}, \C{Add},
and \C{Mul}.
The \C{Neg} tag is associated with an \C{Expr} pointer, while \C{Add} and
\C{Mul} are associated with a \C{BExpr} value.
Since \C{C2RustUnnamed} now contains the tag, the \C{kind} field in the struct
is no longer necessary.

Now, \C{eval} can be implemented with pattern matching as follows:

\vspace{-0.5em}
\begin{lstlisting}[language=Rust,numbers=left]
fn eval(e: *mut Expr) -> i32 {
  match (*e).v {
    C2RustUnnamed::One => return 1,
    C2RustUnnamed::Neg(e) => return -eval(e),
    C2RustUnnamed::Add(b) => return eval(b.l)+eval(b.r),
    C2RustUnnamed::Mul(b) => return eval(b.l)*eval(b.r),
  }}
\end{lstlisting}
\vspace{-0.5em}

\noindent
Each pattern matching branch specifies a tag and binds the associated value to
an identifier.
For instance, in line 4, the associated value is bound to \C{e} and then passed
to \C{eval}.
Since the compiler ensures all variants are covered, the \C{\_} (default) branch
is unnecessary.

The code pattern using \C{if} to check tags is also supported through
\C{if-let}~\cite{rust-if-let}, which is another form of pattern matching.
The following code performs computation only when the tag is \C{Neg}:

\vspace{-0.5em}
\begin{lstlisting}[language=Rust]
if let C2RustUnnamed::Neg(e) = (*e).v {
  return -eval(e); }
\end{lstlisting}
\vspace{-0.5em}

Pattern matching enables the compiler to detect programmers' mistakes that can
cause memory bugs.
Consider the following Rust code where the value associated with \C{Neg} is
treated as a \C{BExpr} type:

\vspace{-0.5em}
\begin{lstlisting}[language=Rust]
match (*e).v {
  C2RustUnnamed::Neg(b) => return eval(b.l) + eval(b.r),
\end{lstlisting}
\vspace{-0.5em}

\noindent
The type of \C{b} is \C{*mut Expr}, as specified in the type definition.
Since \C{*mut Expr} lacks the fields \C{l} and \C{r}, the compiler raises an
error.

Additionally, the compiler verifies the correct construction of tagged union
values.
Consider the following code, which incorrectly initializes a tagged union with
the tag \C{Add} and an \C{Expr} pointer:

\vspace{-0.5em}
\begin{lstlisting}[language=Rust]
let e1: Expr = ...;
let e2 = Expr { v: C2RustUnnamed::Add(&mut e1) };
\end{lstlisting}
\vspace{-0.5em}

\noindent
Since the type definition requires a \C{BExpr} value for \C{Add}, the code does
not pass type checking.

\section{Static Analysis}
\label{sec:analysis}

In this section, we present static analysis designed to facilitate the
transformation of unions with tags into tagged unions.
The objectives of this analysis are to (1) identify a tag field for a union, if
one exists, and (2) determine the tag values associated with each union field.
The proposed static analysis consists of four steps:
\begin{enumerate}
  \item Identification of \emph{candidate structs}, those containing
    unions and their potential tag fields (\cref{sec:analysis:candidate}).
  \item Whole-program may-points-to analysis (\cref{sec:analysis:may}).
  \item Intraprocedural must-points-to analysis for selected functions
    (\cref{sec:analysis:must}).
  \item Interpretation of the analysis results using a heuristic
    (\cref{sec:analysis:heuristic}).
\end{enumerate}
As we analyze Rust code generated by C2Rust, rather than the original C code,
code examples in this section are written in Rust.

\subsection{Candidate Identification}
\label{sec:analysis:candidate}

The first step of the analysis is to identify structs that likely contain unions
and their tag fields.
We also determine which functions to analyze based on the identified structs.
If no such structs are found, the analysis terminates at this step, concluding
that the program does not contain unions to be transformed into tagged unions.

To concretely define candidate structs, we first define \emph{tag-eligible
fields}, those that can potentially serve as tag fields.
A field of a struct is considered tag-eligible if (1) it has an integer type,
(2) when it appears on the left side of an assignment, the operator is \C{=},
excluding others such as \C{+=}, and
(3) it is never referenced by a pointer.
For the first condition, integer types include \C{bool}, \C{i8}, \C{u8},
\C{i16}, \C{u16}, \C{i32}, \C{u32}, \C{i64}, and \C{u64}, which C2Rust
translates from \C{\_Bool}, \C{signed}/\C{unsigned} \C{char}, \C{short},
\C{int}, and \C{long}.
The second and third conditions arise from limitations in the expressibility of
tagged unions.
Tags of tagged unions are not integers and thus cannot undergo integer
operations.
Moreover, since tags do not exist as fields, they cannot be referenced.
Consequently, fields that exhibit such behavior cannot be transformed into
tagged unions.
%

We also define \emph{candidate unions}, which can potentially be accompanied by
tags.
%
A union is a candidate if
(1) it is a field of a struct that contains at least one tag-eligible field, and
(2) its name begins with \C{C2RustUnnamed}.
%
%
The second condition indicates that the union is anonymous in the C code.
If a union has a name, it can be used independently of the struct,
not being expressible as tagged unions.

We finally define candidate structs.
A candidate struct is a struct containing at least one candidate union.

We now describe how to determine which functions to analyze.
To identify tag fields for unions and values associated with union fields, we
need to ascertain the possible values of candidate struct fields.
If a function neither reads from nor writes to a field, intraprocedural analysis
of the function provides no information about that field's value.
Therefore, we analyze only the functions that access fields of candidate
structs.
Such functions can be easily identified through syntactic examination.

\subsection{May-Points-To Analysis}
\label{sec:analysis:may}

The second step of the analysis is to conduct a whole-program may-points-to
analysis.
This step is essential because may-points-to relations are required to ensure
the soundness of the subsequent must-points-to analysis.
The utilization of may-points-to relations in the must-points-to analysis is
detailed in \cref{sec:analysis:must}.

We employ a field-sensitive Andersen-style analysis~\cite{pearce2007efficient}
for the may-points-to analysis.
This analysis is flow-insensitive and has a time complexity of $O(n^3)$.
%
%
Although other may-points-to analyses exist, such as field-insensitive
Andersen-style~\cite{andersen1994program} and Steensgaard-style
analyses~\cite{steensgaard1996pointsa, steensgaard1996points}, they are too
imprecise, leading to unacceptably imprecise must-points-to analysis results.

\subsection{Must-Points-To Analysis}
\label{sec:analysis:must}

The third step is to perform an intraprocedural must-points-to analysis for each
selected function.
Our overall algorithm is akin to typical must-points-to
analyses~\cite{kastrinis2018efficient}.
The execution state at each program point is represented as a graph, with nodes
denoting memory locations and edges expressing points-to relations.
The analysis iteratively updates the state at each program point until reaching
a fixed point.
Each state is derived by joining the state from the previous iteration with the
state resulting from applying the current instruction's effect to the previous
program point's state.

We first describe how we visualize graphs throughout this section.
Since nodes denote memory locations, some nodes correspond to the stack
locations used by local variables.
For clarity in visualization, we draw a dashed arrow from the name of a local variable
to the node representing its memory location.
Note that a variable name is not a node and thus this arrow is not an edge.
%
%
For example, \C{x = \&y} constructs the following graph:

\begin{center}
\begin{tikzpicture}[
  loc/.style={
    circle,
    draw,
    align=center,
    minimum size=0.5cm,
    text width=0.5cm,
    inner sep=0pt
  },
  edge/.style={
    ->,
    >=stealth,
  },
  nameedge/.style={
    ->,
    >=stealth,
    dashed,
    dash pattern=on 1.25pt off 0.75pt,
  },
  scale=0.8,
]
  \node (xname) at (0, 0.4) {\C{x}};
  \node (yname) at (0, 0) {\C{y}};
  \node[loc] (x) at (1, 0.4) {};
  \node[loc] (y) at (2, 0) {};
  \draw[nameedge] (xname) -- (x);
  \draw[nameedge] (yname) -- (y);
  \draw[edge] (x) -- (y);
\end{tikzpicture}
\end{center}

\noindent
The graph has two nodes and one edge and indicates that the pointer at the
memory location of \C{x} points to the memory location of \C{y}.

Since our goal is not only to compute must-points-to relations but also to track
fields' integer values, we optionally label each node with {\sf @$N$}, where $N$
is a set of integers.
The label {\sf @$N$} encodes integer value information in pointer graphs.
If a node is unlabeled, the value at the location is a usual C value (such as an
integer, pointer, struct, union, or array).
If a node is labeled {\sf @$N$}, the value at the location is an imaginary value
at addresses $N$, i.e., the possible addresses of the location are $N$.
Consequently, if a node $v$ has an outgoing edge to a node labeled {\sf @$N$},
then the possible values at $v$'s location are $N$.
For instance, \C{x = 1} constructs the following graph:

\begin{center}
\begin{tikzpicture}[
  loc/.style={
    circle,
    draw,
    align=center,
    minimum size=0.5cm,
    text width=0.5cm,
    inner sep=0pt
  },
  edge/.style={
    ->,
    >=stealth,
  },
  nameedge/.style={
    ->,
    >=stealth,
    dashed,
    dash pattern=on 1.25pt off 0.75pt,
  },
  scale=0.8,
]
  \node (xname) at (0, 0.8) {\C{x}};
  \node[loc] (x) at (1, 0.8) {};
  \node[loc] (xd) at (2, 0.8) {\scriptsize\sf @1};
  \draw[nameedge] (xname) -- (x);
  \draw[edge] (x) -- (xd);
\end{tikzpicture}
\end{center}

Using the {\sf @$N$} label is beneficial since it enables efficient
propagation of integer values to memory locations known to hold the same value.
Consider the following graph, constructed by \C{x = y}:

\begin{center}
\begin{tikzpicture}[
  loc/.style={
    circle,
    draw,
    align=center,
    minimum size=0.5cm,
    text width=0.5cm,
    inner sep=0pt
  },
  edge/.style={
    ->,
    >=stealth,
  },
  nameedge/.style={
    ->,
    >=stealth,
    dashed,
    dash pattern=on 1.25pt off 0.75pt,
  },
  scale=0.8,
]
  \node (xname) at (0, 0.8) {\C{x}};
  \node (yname) at (0, 0) {\C{y}};
  \node[loc] (x) at (1, 0.8) {};
  \node[loc] (y) at (1, 0) {};
  \node[loc] (xd) at (2, 0.4) {};
  \draw[nameedge] (xname) -- (x);
  \draw[edge] (x) -- (xd);
  \draw[nameedge] (yname) -- (y);
  \draw[edge] (y) -- (xd);
\end{tikzpicture}
\end{center}

\noindent
If we obtain the fact that \C{x} equals 1, we update the graph as follows:

\begin{center}
\begin{tikzpicture}[
  loc/.style={
    circle,
    draw,
    align=center,
    minimum size=0.5cm,
    text width=0.5cm,
    inner sep=0pt
  },
  edge/.style={
    ->,
    >=stealth,
  },
  nameedge/.style={
    ->,
    >=stealth,
    dashed,
    dash pattern=on 1.25pt off 0.75pt,
  },
  scale=0.8,
]
  \node (xname) at (0, 0.8) {\C{x}};
  \node (yname) at (0, 0) {\C{y}};
  \node[loc] (x) at (1, 0.8) {};
  \node[loc] (y) at (1, 0) {};
  \node[loc] (xd) at (2, 0.4) {\scriptsize\sf @1};
  \draw[nameedge] (xname) -- (x);
  \draw[edge] (x) -- (xd);
  \draw[nameedge] (yname) -- (y);
  \draw[edge] (y) -- (xd);
\end{tikzpicture}
\end{center}

\noindent
Then, we automatically discover that \C{y} also equals 1.
As this example demonstrates, the {\sf @$N$} label facilitates the update of the
value at multiple memory locations by labeling only a single node.
%
%
%

We now discuss how to analyze code involving unions.
Consider the following example, where \C{Expr} is defined as in
\cref{sec:background:c2rust}:

\vspace{-0.5em}
\begin{lstlisting}[language=Rust, numbers=left]
fn eval(e: *mut Expr) -> i32 {
  let k = (*e).kind;
  match k {
    1 => return -eval((*e).v.e),
\end{lstlisting}
\vspace{-0.5em}

\noindent
After line 2, we have the following graph:

\begin{center}
\begin{tikzpicture}[
  loc/.style={
    circle,
    draw,
    align=center,
    minimum size=0.5cm,
    text width=0.5cm,
    inner sep=0pt
  },
  edge/.style={
    ->,
    >=stealth,
  },
  nameedge/.style={
    ->,
    >=stealth,
    dashed,
    dash pattern=on 1.25pt off 0.75pt,
  },
  scale=0.8,
]
  \node (ename) at (0, 0.8) {\C{e}};
  \node (kindname) at (0, 0) {\C{k}};
  \node[loc] (e) at (1, 0.8) {};
  \node[loc] (kind) at (1, 0) {};
  \node[loc] (ed) at (2, 0.8) {};
  \node[loc] (edd) at (3.5, 0.4) {};
  \draw[nameedge] (ename) -- (e);
  \draw[edge] (e) -- (ed);
  \draw[edge] (ed) -- (edd) node[midway, above] {\scriptsize\tt .kind};
  \draw[nameedge] (kindname) -- (kind);
  \draw[edge] (kind) -- (edd);
\end{tikzpicture}
\end{center}

\noindent
Here, the edge labeled \C{.kind} indicates the presence of a struct/union at the
location of the node where the edge originates, with the pointer stored in the
\C{kind} field referring to the pointed node's location.
Since line 3 uses \C{k} as the condition for \C{match}, we determine that \C{k}
equals \C{1} in line 4.
Consequently, the graph at the beginning of line 4 is as follows:

\begin{center}
\begin{tikzpicture}[
  loc/.style={
    circle,
    draw,
    align=center,
    minimum size=0.5cm,
    text width=0.5cm,
    inner sep=0pt
  },
  edge/.style={
    ->,
    >=stealth,
  },
  nameedge/.style={
    ->,
    >=stealth,
    dashed,
    dash pattern=on 1.25pt off 0.75pt,
  },
  scale=0.8,
]
  \node (ename) at (0, 0.8) {\C{e}};
  \node (kindname) at (0, 0) {\C{k}};
  \node[loc] (e) at (1, 0.8) {};
  \node[loc] (kind) at (1, 0) {};
  \node[loc] (ed) at (2, 0.8) {};
  \node[loc] (edd) at (3.5, 0.4) {\scriptsize\sf @1};
  \draw[nameedge] (ename) -- (e);
  \draw[edge] (e) -- (ed);
  \draw[edge] (ed) -- (edd) node[midway, above] {\scriptsize\tt .kind};
  \draw[nameedge] (kindname) -- (kind);
  \draw[edge] (kind) -- (edd);
\end{tikzpicture}
\end{center}

\noindent
From this graph, we conclude that the value of \C{kind} is \C{1} when the union
field \C{e} is accessed in line 4.

More complex scenarios involve joining two graphs.
Consider the following code example:

\vspace{-0.5em}
\begin{lstlisting}[language=Rust, numbers=left]
match (*e).kind {
  2 => ...,
  3 => ...,
  _ => return, }
let lv = eval((*e).v.b.l);
\end{lstlisting}
\vspace{-0.5em}

\noindent
In lines 2 and 3, we have the following graphs, respectively:

\begin{center}
\begin{tikzpicture}[
  loc/.style={
    circle,
    draw,
    align=center,
    minimum size=0.5cm,
    text width=0.5cm,
    inner sep=0pt
  },
  edge/.style={
    ->,
    >=stealth,
  },
  nameedge/.style={
    ->,
    >=stealth,
    dashed,
    dash pattern=on 1.25pt off 0.75pt,
  },
  scale=0.8,
]
  \node (ename) at (0, 0.8) {\C{e}};
  \node[loc] (e) at (1, 0.8) {};
  \node[loc] (ed) at (2, 0.8) {};
  \node[loc] (edd) at (3.5, 0.8) {\scriptsize\sf @2};
  \draw[nameedge] (ename) -- (e);
  \draw[edge] (e) -- (ed);
  \draw[edge] (ed) -- (edd) node[midway, above] {\scriptsize\tt .kind};

  \node (ename2) at (5, 0.8) {\C{e}};
  \node[loc] (e2) at (6, 0.8) {};
  \node[loc] (ed2) at (7, 0.8) {};
  \node[loc] (edd2) at (8.5, 0.8) {\scriptsize\sf @3};
  \draw[nameedge] (ename2) -- (e2);
  \draw[edge] (e2) -- (ed2);
  \draw[edge] (ed2) -- (edd2) node[midway, above] {\scriptsize\tt .kind};
\end{tikzpicture}
\end{center}

\noindent
Since line 5 is reachable from both lines 2 and 3, we need to join the graphs.
When joining graphs, the integer sets in the labels are unioned, resulting in the
following graph:

\begin{center}
\begin{tikzpicture}[
  loc/.style={
    circle,
    draw,
    align=center,
    minimum size=0.5cm,
    text width=0.5cm,
    inner sep=0pt
  },
  edge/.style={
    ->,
    >=stealth,
  },
  nameedge/.style={
    ->,
    >=stealth,
    dashed,
    dash pattern=on 1.25pt off 0.75pt,
  },
  scale=0.8,
]
  \node (ename) at (0, 0.8) {\C{e}};
  \node[loc] (e) at (1, 0.8) {};
  \node[loc] (ed) at (2, 0.8) {};
  \node[loc] (edd) at (3.5, 0.8) {\scriptsize\sf @2,3};
  \draw[nameedge] (ename) -- (e);
  \draw[edge] (e) -- (ed);
  \draw[edge] (ed) -- (edd) node[midway, above] {\scriptsize\tt .kind};
\end{tikzpicture}
\end{center}

\noindent
From the graph, we conclude that the value of \C{kind} is \C{2} or \C{3} when
the union field \C{b} is accessed in line 5.

The join of graphs is carefully defined to maintain the soundness of the
analysis.
Consider the following code, where control flow splits based on the value of
\C{kind} and then merges:

\vspace{-0.5em}
\begin{lstlisting}[language=Rust, numbers=left]
if (*e).kind == 1 { ... }
else { ... }
...
\end{lstlisting}
\vspace{-0.5em}

\noindent
The states in lines 1 and 2 are as follows, respectively:

\begin{center}
\begin{tikzpicture}[
  loc/.style={
    circle,
    draw,
    align=center,
    minimum size=0.5cm,
    text width=0.5cm,
    inner sep=0pt
  },
  edge/.style={
    ->,
    >=stealth,
  },
  nameedge/.style={
    ->,
    >=stealth,
    dashed,
    dash pattern=on 1.25pt off 0.75pt,
  },
  scale=0.8,
]
  \node (ename) at (0, 0.8) {\C{e}};
  \node[loc] (e) at (1, 0.8) {};
  \node[loc] (ed) at (2, 0.8) {};
  \node[loc] (edd) at (3.5, 0.8) {\scriptsize\sf @1};
  \draw[nameedge] (ename) -- (e);
  \draw[edge] (e) -- (ed);
  \draw[edge] (ed) -- (edd) node[midway, above] {\scriptsize\tt .kind};

  \node (ename2) at (5, 0.8) {\C{e}};
  \node[loc] (e2) at (6, 0.8) {};
  \node[loc] (ed2) at (7, 0.8) {};
  \node[loc] (edd2) at (8.5, 0.8) {};
  \draw[nameedge] (ename2) -- (e2);
  \draw[edge] (e2) -- (ed2);
  \draw[edge] (ed2) -- (edd2) node[midway, above] {\scriptsize\tt .kind};
\end{tikzpicture}
\end{center}

\noindent
In line 1, \C{(*e).kind} is known to be \C{1}.
However, in line 2, its value is unknown.
When entering line 3, these graphs are joined as follows:

\begin{center}
\begin{tikzpicture}[
  loc/.style={
    circle,
    draw,
    align=center,
    minimum size=0.5cm,
    text width=0.5cm,
    inner sep=0pt
  },
  edge/.style={
    ->,
    >=stealth,
  },
  nameedge/.style={
    ->,
    >=stealth,
    dashed,
    dash pattern=on 1.25pt off 0.75pt,
  },
  scale=0.8,
]
  \node (ename) at (0, 0.8) {\C{e}};
  \node[loc] (e) at (1, 0.8) {};
  \node[loc] (ed) at (2, 0.8) {};
  \node[loc] (edd) at (3.5, 0.8) {};
  \draw[nameedge] (ename) -- (e);
  \draw[edge] (e) -- (ed);
  \draw[edge] (ed) -- (edd) node[midway, above] {\scriptsize\tt .kind};
\end{tikzpicture}
\end{center}

\noindent
Since the node from line 2 is unlabeled, the joined graph's node is also
unlabeled.
This indicates that we do not know the value of \C{(*e).kind}, which
is correct.
This example shows that no-label signifies no-information regarding the address
of the node's location, i.e., it represents ${\sf @}\mathbb{Z}$ (all integers),
not ${\sf @}\emptyset$ (empty set).

\SetStartEndCondition{ }{}{}
\SetKwProg{Fn}{def}{\string:}{}
\SetKwFunction{Range}{range}
\SetKw{KwTo}{in}\SetKwFor{For}{for}{\string:}{}
\SetKwIF{If}{ElseIf}{Else}{if}{:}{elif}{else:}{}
\SetKwFor{While}{while}{:}{fintq}
\AlgoDontDisplayBlockMarkers\SetAlgoNoEnd\SetAlgoNoLine

\SetKwInOut{Input}{Input}
\SetKwInOut{Output}{Output}

\begin{algorithm}[t]
  \caption{Graph joining}
  \label{alg:join}
  \setstretch{0.9}
  \small
  \Input{$g_1$, $g_2$}
  \Output{$g$}
  $g.\mathsf{nodes}:=\emptyset$;\ $g.\mathsf{edges}:=\emptyset$;\ $\mathsf{worklist}:=\emptyset$\;
  \For{$x\leftarrow\text{function's local variables}$}{
    \If{$g_1\ \text{has a node}\ v_1\ \text{corresponding to}\ x$}{
      \If{$g_2\ \text{has a node}\ v_2\ \text{corresponding to}\ x$}{
        $g.\mathsf{nodes}.\mathsf{insert}((v_1,v_2))$\;
        $\mathsf{worklist}.\mathsf{insert}((v_1,v_2))$\;
      }
    }
  }
  \While{$\mathsf{worklist}\not=\emptyset$}{
    $(v_1,v_2):=\mathsf{worklist}.\mathsf{pop}()$\;
    \For{$(v_1,v_1',f)\leftarrow v_1\text{'s outgoing edges in}\ g_1$}{
      \If{$g_2\ \text{has an edge}\ (v_2,v_2',f)$}{
        $g.\mathsf{edges}.\mathsf{insert}((v_1,v_2),(v_1',v_2'),f)$\;
        \If{$(v_1',v_2')\not\in g.\mathsf{nodes}$}{
          $g.\mathsf{nodes}.\mathsf{insert}((v_1',v_2'))$\;
          $\mathsf{worklist}.\mathsf{insert}((v_1',v_2'))$\;
        }
      }
    }
    \If{$v_1\ \text{has a label}\ {\sf @}N_1\ \text{in}\ g_1$}{
      \If{$v_2\ \text{has a label}\ {\sf @}N_2\ \text{in}\ g_2$}{
        $\text{set}\ (v_1,v_2)\text{'s label to}\ {\sf @}(N_1\cup N_2)\ \text{in}\ g$\;
      }
    }
  }
\end{algorithm}

\cref{alg:join} illustrates the algorithm for graph joining.
Edges are intersected to retain must-point-to relations, while integer sets in
labels are unioned.
In the pseudo-code, each node in the input graphs $g_1$ and $g_2$ is denoted by
a unique identifier $v$, and each node in the output graph $g$ is represented by
a pair of identifiers $(v_1,v_2)$.
Each edge is represented by $(v,v',f)$, denoting an edge labeled $f$ from node
$v$ to node $v'$.
Unlabeled edges are treated as edges with the empty label $\epsilon$.
Initially, $g$ and the worklist, which stores $g$'s nodes to be visited, are
both empty (line 1).
Then, we add nodes corresponding to local variables to $g$ (lines 2--6).
Finally, we visit each added node until the worklist is empty (lines 7--8).
During the visit, edges with the same label in $g_1$ and $g_2$ are added to $g$
(lines 9--14), and node labels are unioned if they exist (lines 15--17).

We now discuss the utilization of may-points-to relations during the must-points-to analysis.
Consider the following code:

\vspace{-0.5em}
\begin{lstlisting}[language=Rust, numbers=left]
let e = if ... { &mut ev } else { ... };
let k = (*e).kind;
ev = ...;
match k {
  1 => ...,
\end{lstlisting}
\vspace{-0.5em}

\noindent
Since \C{e} may not point to \C{ev}, the state after line 1 is as follows:

\begin{center}
\begin{tikzpicture}[
  loc/.style={
    circle,
    draw,
    align=center,
    minimum size=0.5cm,
    text width=0.5cm,
    inner sep=0pt
  },
  edge/.style={
    ->,
    >=stealth,
  },
  nameedge/.style={
    ->,
    >=stealth,
    dashed,
    dash pattern=on 1.25pt off 0.75pt,
  },
  scale=0.8,
]
  \node (ename) at (0, 0.8) {\C{e}};
  \node (evname) at (3, 0.8) {\C{ev}};
  \node[loc] (e) at (1, 0.8) {};
  \node[loc] (ev) at (4, 0.8) {};
  \node[loc] (ed) at (2, 0.8) {};
  \draw[nameedge] (ename) -- (e);
  \draw[edge] (e) -- (ed);
  \draw[nameedge] (evname) -- (ev);
\end{tikzpicture}
\end{center}

\noindent
This graph indicates that \C{e} points to some location, not necessarily the
same as \C{ev}'s location.
Line 2 updates the graph as follows:

\begin{center}
\begin{tikzpicture}[
  loc/.style={
    circle,
    draw,
    align=center,
    minimum size=0.5cm,
    text width=0.5cm,
    inner sep=0pt
  },
  edge/.style={
    ->,
    >=stealth,
  },
  nameedge/.style={
    ->,
    >=stealth,
    dashed,
    dash pattern=on 1.25pt off 0.75pt,
  },
  scale=0.8,
]
  \node (ename) at (0, 0.8) {\C{e}};
  \node (kindname) at (0, 0) {\C{k}};
  \node[loc] (e) at (1, 0.8) {};
  \node[loc] (kind) at (1, 0) {};
  \node[loc] (ed) at (2, 0.8) {};
  \node[loc] (edd) at (3.5, 0.4) {};
  \draw[nameedge] (ename) -- (e);
  \draw[edge] (e) -- (ed);
  \draw[edge] (ed) -- (edd) node[midway, above] {\scriptsize\tt .kind};
  \draw[nameedge] (kindname) -- (kind);
  \draw[edge] (kind) -- (edd);

  \node (evname) at (4.5, 0.4) {\C{ev}};
  \node[loc] (ev) at (5.5, 0.4) {};
  \draw[nameedge] (evname) -- (ev);
\end{tikzpicture}
\end{center}

\noindent
Line 3 mutates the value of \C{ev}, possibly changing the value of
\C{(*e).kind}.
Thus, \C{(*e).kind} is not necessarily equal to \C{k} after line 3,
necessitating an appropriate update to the graph.
As the graph itself does not reveal any relations between \C{e} and \C{ev}, we
rely on the precomputed may-points-to relations, which indicate that \C{e} may
point to \C{ev}.
Consequently, we remove all outgoing edges, including the \C{kind} edge, from
\C{*e}'s node, resulting in the following graph:

\begin{center}
\begin{tikzpicture}[
  loc/.style={
    circle,
    draw,
    align=center,
    minimum size=0.5cm,
    text width=0.5cm,
    inner sep=0pt
  },
  edge/.style={
    ->,
    >=stealth,
  },
  nameedge/.style={
    ->,
    >=stealth,
    dashed,
    dash pattern=on 1.25pt off 0.75pt,
  },
  scale=0.8,
]
  \node (ename) at (0, 0.8) {\C{e}};
  \node (kindname) at (0, 0) {\C{k}};
  \node[loc] (e) at (1, 0.8) {};
  \node[loc] (kind) at (1, 0) {};
  \node[loc] (ed) at (2, 0.8) {};
  \node[loc] (edd) at (3.5, 0.4) {};
  \draw[nameedge] (ename) -- (e);
  \draw[edge] (e) -- (ed);
  \draw[nameedge] (kindname) -- (kind);
  \draw[edge] (kind) -- (edd);

  \node (evname) at (4.5, 0.4) {\C{ev}};
  \node[loc] (ev) at (5.5, 0.4) {};
  \draw[nameedge] (evname) -- (ev);
\end{tikzpicture}
\end{center}

\noindent
In line 5, we successfully avoid the incorrect conclusion that \C{(*e).kind}
equals \C{1}, as shown in the following graph:

\begin{center}
\begin{tikzpicture}[
  loc/.style={
    circle,
    draw,
    align=center,
    minimum size=0.5cm,
    text width=0.5cm,
    inner sep=0pt
  },
  edge/.style={
    ->,
    >=stealth,
  },
  nameedge/.style={
    ->,
    >=stealth,
    dashed,
    dash pattern=on 1.25pt off 0.75pt,
  },
  scale=0.8,
]
  \node (ename) at (0, 0.8) {\C{e}};
  \node (kindname) at (0, 0) {\C{k}};
  \node[loc] (e) at (1, 0.8) {};
  \node[loc] (kind) at (1, 0) {};
  \node[loc] (ed) at (2, 0.8) {};
  \node[loc] (edd) at (3.5, 0.4) {\scriptsize\sf @1};
  \draw[nameedge] (ename) -- (e);
  \draw[edge] (e) -- (ed);
  \draw[nameedge] (kindname) -- (kind);
  \draw[edge] (kind) -- (edd);

  \node (evname) at (4.5, 0.4) {\C{ev}};
  \node[loc] (ev) at (5.5, 0.4) {};
  \draw[nameedge] (evname) -- (ev);
\end{tikzpicture}
\end{center}

\noindent
Like this, the analysis removes the appropriate edges from the graph at each
(indirect) assignment and function call according to the may-points-to
relations.
The effect of a function call is equivalent to the cumulative effects of all
assignments reachable by the call.

\subsection{Analysis Result Interpretation}
\label{sec:analysis:heuristic}

\SetKwFunction{Fcacc}{CollectFromAccesses}
\SetKwFunction{Fcstr}{CollectFromStructs}
\SetKwFunction{Fcall}{CollectAllTags}
\SetKwFunction{Fidentify}{IdentifyTags}

\begin{algorithm}[t]
  \caption{Identifying tag values associated with fields}
  \label{alg:heuristic}
  \setstretch{0.9}
  \small
  \Fn{\Fidentify{struct $s$, union $u$, field $f_s$}}{
    $res:=$ \Fcacc{$s,u,f_s$}\;
    \If{$res=\mathsf{None}$}{
      \Return{$\mathsf{None}$}\;
    }
    $(\mathsf{field\_tags},\mathsf{access\_tags}):=res$\;
    $\mathsf{field\_tags}':=$ \Fcstr{$s,u,f_s$}\;
    $\mathsf{struct\_tags}:=\emptyset$\;
    \For{$f_u\leftarrow\text{fields of}\ u$}{
      $\mathsf{tags}:=\mathsf{field\_tags'}[f_u]\setminus\mathsf{access\_tags}$\;
      \If{$\mathsf{struct\_tags}\cap\mathsf{tags}\not=\emptyset$}{
        \Return{$\mathsf{None}$}\;
      }
      $\mathsf{field\_tags}[f_u]:=\mathsf{field\_tags}[f_u]\cup\mathsf{tags}$\;
      $\mathsf{struct\_tags}:=\mathsf{struct\_tags}\cup\mathsf{tags}$\;
    }
    $\mathsf{all\_tags}:=$ \Fcall{$s,u,f_s$}\;
    $\mathsf{rem\_tags}:=\mathsf{all\_tags}\setminus(\mathsf{access\_tags}\cup\mathsf{struct\_tags})$\;
    \Return{$(\mathsf{field\_tags},\mathsf{rem\_tags})$}\;
  }
  \Fn{\Fcacc{struct $s$, union $u$, field $f_s$}}{
    $\mathsf{field\_tags}:=\mathsf{Map}()$; $\mathsf{all\_tags}:=\emptyset$\;
    \For{$l\leftarrow \text{analyzed program points}$}{
      \If{$\text{a field}\ f_u\ \text{of}\ u\ \text{is accessed}\ \text{at}\ l$}{
        $N:=\text{possible values of}\ f_s\ \text{at}\ l$\;
        \If{$N\ \text{is from}\ \C{if}\ or\ \C{match}$}{
          $\mathsf{tags}:=N\setminus\mathsf{field\_tags}[f_u]$\;
          \If{$\mathsf{all\_tags}\cap\mathsf{tags}\not=\emptyset$}{
            \Return{$\mathsf{None}$}\;
          }
          $\mathsf{field\_tags}[f_u]:=\mathsf{field\_tags}[f_u]\cup \mathsf{tags}$\;
          $\mathsf{all\_tags}:=\mathsf{all\_tags}\cup \mathsf{tags}$\;
        }
      }
    }
    \Return{$(\mathsf{field\_tags},\mathsf{all\_tags})$}\;
  }
  \Fn{\Fcstr{struct $s$, union $u$, field $f_s$}}{
    $\mathsf{field\_tags}:=\mathsf{Map}()$\;
    \For{$l\leftarrow \text{analyzed program points}$}{
      \If{$l\ \text{is end of a basic block}$}{
        \For{$v\leftarrow \text{struct}\ s\ \text{reachable at}\ l$}{
          \If{$\text{union}\ u\ \text{in}\ v\ \text{has a field}\ f_u$}{
            $N:=\text{possible values of}\ f_s\ \text{of}\ v$\;
            $\mathsf{field\_tags}[f_u]:=\mathsf{field\_tags}[f_u]\cup N$\;
          }
        }
      }
    }
    \Return{$\mathsf{field\_tags}$}\;
  }
  \Fn{\Fcall{struct $s$, union $u$, field $f_s$}}{
    $\mathsf{all\_tags}:=\emptyset$\;
    \For{$l\leftarrow \text{analyzed program points}$}{
      \For{$v\leftarrow \text{struct}\ s\ \text{reachable at}\ l$}{
        $N:=\text{possible values of}\ f_s\ \text{of}\ v$\;
        $\mathsf{all\_tags}:=\mathsf{all\_tags}\cup N$\;
      }
    }
    \Return{$\mathsf{all\_tags}$}\;
  }
\end{algorithm}

The last step of the analysis is to interpret the results of the must-points-to
analysis using a heuristic, as demonstrated in \cref{alg:heuristic}.
The entry point is \C{IdentifyTags}, which takes three arguments: a candidate
struct $s$, a candidate union $u$ in $s$, and a tag-eligible field $f_s$ in $s$.
Its goal is to determine whether $f_s$ serves as a tag field for $u$ and, if it
does, to identify the tag values associated with each field of $u $.

As the first step of the heuristic, we invoke \C{CollectFromAccesses} (line 1)
to identify the associated tag values by examining the value of $f_s$ when a
union field is accessed after $f_s$ has been checked by \C{match}/\C{if}.
This subroutine returns $\mathsf{field\_tags}$, a map from union fields to their
tag values, and $\mathsf{all\_tags}$, a set containing all tag values in
$\mathsf{field\_tags}$.
Initially, both are empty (line 18).
We then iterate over every program point in the analyzed functions and check if
any field of $u$ is accessed (lines 19--20).
If accessed, we determine the possible values of $f_s$ from the {\sf @$N$} label
of the graph computed by the must-points-to analysis (line 21).
Here, we treat no-label as the empty set.
The possible values should originate from \C{match}/\C{if} on $f_s$, and not
from an assignment to $f_s$ (line 22).
If any of these values are already associated with other fields, we immediately
return $\mathsf{None}$, indicating that $f_s$ is not a tag field (lines 23--25).
Otherwise, we add the values to $\mathsf{field\_tags}$ and $\mathsf{all\_tags}$
(lines 26--27).

If \C{CollectFromAccesses} succeeds, \C{IdentifyTags} proceeds to the next step
by calling \C{CollectFromStructs} (line 6).
This complements the previous step by discovering tag-field associations that
may have been missed due to the analysis's imprecision.
This subroutine considers the field values and the last-written union fields at
the end of each basic block.
For instance, consider the following code:

\vspace{-0.5em}
\begin{lstlisting}[language=Rust]
(*e).kind = 2; (*e).v.b = ...; return e;
\end{lstlisting}
\vspace{-0.5em}

\noindent
We have the following graph at return:

\begin{center}
\begin{tikzpicture}[
  loc/.style={
    circle,
    draw,
    align=center,
    minimum size=0.5cm,
    text width=0.5cm,
    inner sep=0pt
  },
  edge/.style={
    ->,
    >=stealth,
  },
  nameedge/.style={
    ->,
    >=stealth,
    dashed,
    dash pattern=on 1.25pt off 0.75pt,
  },
  scale=0.8,
]
  \node (ename) at (0, 0.8) {\C{e}};
  \node[loc] (e) at (1, 0.8) {};
  \node[loc] (ed) at (2, 0.8) {};
  \node[loc] (edd) at (3.5, 1.2) {\scriptsize\sf @2};
  \node[loc] (eddv) at (3.5, 0.4) {};
  \draw[nameedge] (ename) -- (e);
  \draw[edge] (e) -- (ed);
  \draw[edge] (ed) -- (edd) node[midway, above] {\scriptsize\tt .kind};
  \draw[edge] (ed) -- (eddv) node[midway, above] {\scriptsize\tt .v.b};
\end{tikzpicture}
\end{center}

\noindent
From this, we can conclude that \C{kind} equals \C{2} when \C{b} is the
last-written union field, likely associating \C{2} with \C{b}.

We inspect only the states at the end of basic blocks because the states of
other program points are prone to provide incorrect information from
\emph{intermediate} states.
Consider the following example:

\vspace{-0.5em}
\begin{lstlisting}[language=Rust, numbers=left]
(*e).kind = 1; (*e).v.e = ...; ...
(*e).kind = 2;
(*e).v.b = ...; return e;
\end{lstlisting}
\vspace{-0.5em}

\noindent
Initially, \C{e} is used as a negation expression by setting \C{kind} to \C{1},
but it becomes an addition expression by setting \C{kind} to \C{2} in the end.
If we examine the state after line 2, we will get an incorrect association
between \C{2} and \C{e} from the following graph:

\begin{center}
\begin{tikzpicture}[
  loc/.style={
    circle,
    draw,
    align=center,
    minimum size=0.5cm,
    text width=0.5cm,
    inner sep=0pt
  },
  edge/.style={
    ->,
    >=stealth,
  },
  nameedge/.style={
    ->,
    >=stealth,
    dashed,
    dash pattern=on 1.25pt off 0.75pt,
  },
  scale=0.8,
]
  \node (ename) at (0, 0.8) {\C{e}};
  \node[loc] (e) at (1, 0.8) {};
  \node[loc] (ed) at (2, 0.8) {};
  \node[loc] (edd) at (3.5, 1.2) {\scriptsize\sf @2};
  \node[loc] (eddv) at (3.5, 0.4) {};
  \draw[nameedge] (ename) -- (e);
  \draw[edge] (e) -- (ed);
  \draw[edge] (ed) -- (edd) node[midway, above] {\scriptsize\tt .kind};
  \draw[edge] (ed) -- (eddv) node[midway, above] {\scriptsize\tt .v.e};
\end{tikzpicture}
\end{center}

\noindent
To avoid this issue, we examine only the end of basic blocks, where both the tag
value and the union field are likely to be correctly set.

However, this approach cannot completely prevent reading intermediate states.
For example, a function call can occur between the tag set and the union field
set.
Therefore, we prioritize the data from \C{CollectFromAccesses}, which are less
likely to be affected by intermediate states, over \C{CollectFromStructs}.
If \C{CollectFromAccesses} associates a certain tag value with a specific union
field, that tag value in \C{CollectFromStructs}' results is ignored.

To implement this approach, \C{CollectFromStructs} iterates over the end of
every basic block in each analyzed function, identifying every value of type $s$
reachable from local variables at that point (lines 31--33).
If the struct's union has a written field, that field is associated with the
values of $f_s$ (lines 34--36).
Then, \C{IdentifyTags} combines results from both subroutines, prioritizing data
from \C{CollectFromAccesses} (lines 8--13).
It removes tags already associated with union fields by \C{CollectFromAccesses}
from the results of \C{CollectFromStructs} (line 9) and returns $\mathsf{None}$
if a tag value remains associated with two fields even after this removal (lines
10--11).

Finally, we search for tag values not associated with any union fields.
To achieve this, we call \C{CollectAllTags} (line 14), which collects all
possible tag values by examining the state at every program point (lines
40--43).
By removing the tags associated with union fields, we isolate the tags not
associated with union fields (line 15).

To determine the tag field for each union, we run \C{IdentifyTags} on all
tag-eligible fields.
The field for which \C{IdentifyTags} returns a value other than $\mathsf{None}$
is identified as the tag field.
If multiple tag fields are found, we select the one with the highest number of
distinct tag values.

\section{Code Transformation}
\label{sec:transformation}

In this section, we present code transformation that replaces unions with tagged
unions in C2Rust-generated code, using the results of the static analysis.
We demonstrate the transformation of the \C{Expr} type defined in
\cref{sec:background:c2rust} as an example.
We assume that the tag field and the tag values associated with each union field
are correctly identified by the static analysis.

The type definitions are transformed as follows:

\vspace{-0.5em}
\begin{lstlisting}[language=Rust]
struct Expr { v: C2RustUnnamed }
enum C2RustUnnamed {
  Empty0, e1(*mut Expr), b2(BExpr), b3(BExpr) }
struct BExpr { l: *mut Expr, r: *mut Expr }
\end{lstlisting}
\vspace{-0.5em}

\noindent
This result is the same as the hand-written code in
\cref{sec:background:tagged-union}, except for the variant names.
We generate variant names by concatenating the union field name with the tag
value.
For tags not associated with any fields, we prepend \C{Empty} to the tag value.
To improve variant names, one option is to utilize global variables' names.
When using unions with tag fields, C programmers often define an \emph{enum},
i.e., a group of constant integers with associated names, to use them as tag
values.
These names typically reflect the programmers' understanding of the tag values'
meaning, e.g., \C{EXPR\_ONE}.
Since C2Rust translates each enum definition into multiple constant global
variable definitions while preserving the names, we can use these variable names
instead of the union field names.
Although the names generated with this strategy may still be unsatisfactory,
developers can easily rename them to more meaningful names using IDEs.

We now discuss the transformation of code using unions.
We propose two approaches: na\"ive transformation, which can be applied to any
code but does not adhere to Rust idioms (\cref{sec:transformation:naive}), and
idiomatic transformation, which follows Rust idioms but is applicable only to
specific code patterns (\cref{sec:transformation:idiomatic}).
We use both methods within a single codebase, prioritizing idiomatic
transformation wherever possible and resorting to na\"ive transformation when
necessary.

\subsection{Na\"ive Transformation}
\label{sec:transformation:naive}

Naïve transformation involves defining helper methods for the transformed
structs and unions.
These methods are categorized into two groups: reading and writing.
We first focus on reading. Below are the read-related methods for \C{Expr} and
\C{C2RustUnnamed}:

\vspace{-0.5em}
\begin{lstlisting}[language=Rust,numbers=left]
impl Expr {
  fn kind(self) -> i32 {
    match self.v {
      C2RustUnnamed::Empty0 => 0,
      C2RustUnnamed::e1(_) => 1, ... }}}
impl C2RustUnnamed {
  fn get_e(self) -> *mut Expr {
    if let C2RustUnnamed::e1(v) = self { v }
    else { panic!() }}}
\end{lstlisting}
\vspace{-0.5em}

The \C{kind} method (lines 2--5) of \C{Expr} replaces the \C{kind} field in the
original code.
This method returns the appropriate tag value by applying pattern matching to
the tagged union value.
Code that reads a tag field is replaced with a call to the tag-returning method.

The \C{get\_e} method (lines 7--9) of \C{C2RustUnnamed} replaces the union field
\C{e}.
This method returns a value by applying pattern matching to the tagged union
value.
If the current variant does not contain such a value, the method triggers a
panic.
This allows the dynamic detection of potential bugs by identifying read from a
field other than the last-written one, rather than silently reinterpreting the
value.
Although not shown in the example, a method \C{get\_}$f$ is defined similarly
for each union field $f$.
We replace code reading a union field with a call to the corresponding getter
method.

Using these methods, we transform code reading tags and union fields as follows,
where the former represents the code before transformation and the latter
represents the code after transformation:

\vspace{-0.5em}
\begin{lstlisting}[language=Rust]
match (*e).kind { 1 => eval((*e).v.e),         // before
\end{lstlisting}
\vspace{-1em}
\begin{lstlisting}[language=Rust]
match (*e).kind() { 1 => eval((*e).v.get_e()), // after
\end{lstlisting}
\vspace{-0.5em}

\noindent
Although this transformation preserves the semantics, the resulting code is not
idiomatic.
It applies pattern matching twice to the tagged union value---once to get the
tag value and once to get the union field value---instead of applying pattern
matching only once to directly access the associated value of each variant.

We now describe the write-related methods, defined as follows:

\vspace{-0.5em}
\begin{lstlisting}[language=Rust,numbers=left]
impl Expr {
  fn set_kind(&mut self, v: i32) {
    match v {
      0 => { self.v = C2RustUnnamed::Empty0 }
      1 => {
        let v =
          if let C2RustUnnamed::e1(v) = self.v { v }
          else { std::ptr::null_mut() };
        self.v = C2RustUnnamed::e1(v) } ... }}}
impl C2RustUnnamed {
  fn deref_e_mut(&mut self) -> *mut *mut Expr {
      if let C2RustUnnamed::e1(_) = self {}
      else { *self = Self::e1(std::ptr::null_mut()); }
      if let C2RustUnnamed::e1(v) = self { v }
      else { panic!() }}}
\end{lstlisting}
\vspace{-0.5em}

The \C{set\_kind} method (lines 2--9) of \C{Expr} replaces assignments to
\C{kind}.
It takes a tag value as an argument and updates the tagged union to the
appropriate variant.
If the variant has an associated value, we check if this value already exists
and reuse it if it does (line 7).
If the value does not exist, we create an arbitrary value, which in this case is
the null pointer (line 8).

The \C{deref\_e\_mut} method (lines 11--15) of \C{C2RustUnnamed} provides a
pointer to the inner value, which we use to replace code that mutates \C{e} or
takes its address.
First, we check whether the current variant is appropriate (line 12), and, if
not, update the value to the correct variant (line 13).
Then, we return a pointer to the value (line 14), while the \C{panic!()} on line
15 is never reached.
We also define a method \C{deref\_}$f$\C{\_mut} for each union field $f$ in a
similar manner.

We transform code that updates tags and union fields as follows:

\vspace{-0.5em}
\begin{lstlisting}[language=Rust]
(*e).kind = 1; (*e).v.e = ...;                 // before
\end{lstlisting}
\vspace{-1em}
\begin{lstlisting}[language=Rust]
(*e).set_kind(1); *(*e).v.deref_e_mut() = ...; // after
\end{lstlisting}
\vspace{-0.5em}

\noindent
In the transformed code, \C{set\_kind} changes the variant to \C{e1}.
Then, \C{deref\_e\_mut} returns a pointer to the arbitrary value set by
\C{set\_kind}, and the indirect assignment to the pointer updates this value.

Note that \C{deref\_e\_mut} does not trigger a panic even when the current
variant differs from the expected one, unlike \C{get\_e}.
This behavior ensures the correct transformation of code that first writes to a
union field and then sets the tag.
For example, the following transformation preserves the semantics:

\vspace{-0.5em}
\begin{lstlisting}[language=Rust]
(*e).v.e = ...; (*e).kind = 1;                 // before
\end{lstlisting}
\vspace{-1em}
\begin{lstlisting}[language=Rust]
*(*e).v.deref_e_mut() = ...; (*e).set_kind(1); // after
\end{lstlisting}
\vspace{-0.5em}

\noindent
After the transformation, \C{deref\_e\_mut} changes the variant to \C{e1}, and
the indirect assignment to the pointer sets the associated value. Then,
\C{set\_kind} retains both the variant and associated value.
Although this approach preserves the semantics, it is not idiomatic in Rust, as
we can create a tagged union value within a single expression instead of setting
the tag and the union field separately.

\subsection{Idiomatic Transformation}
\label{sec:transformation:idiomatic}

Idiomatic transformation uses pattern matching on tagged union values and
constructs a tagged union value with a single expression, avoiding helper
methods.
Below is the transformation of \C{match}:

\vspace{-0.5em}
\begin{lstlisting}[language=Rust]
match (*e).kind { 1 => eval((*e).v.e),         // before
\end{lstlisting}
\vspace{-1em}
\begin{lstlisting}[language=Rust]
                                               // after
match (*e).v { C2RustUnnamed::e1(ref x) => eval(*x),
\end{lstlisting}
\vspace{-0.5em}

\noindent
The \C{match} condition is the tagged union value itself, rather than the tag
value obtained by \C{get\_kind}.
In addition, each branch directly matches the variant, instead of comparing the
tag value to an integer.
The \C{ref} keyword before the identifier \C{x} binds a pointer to the
associated value, not the value itself, to \C{x}, allowing the branch to read
and modify the value.
This eliminates the need to call methods such as \C{get\_e} and
\C{deref\_e\_mut}, as \C{x} directly accesses the value.

We also transform \C{if} that checks tag values to use pattern matching with
\C{if-let}, as illustrated in the following example:

\vspace{-0.5em}
\begin{lstlisting}[language=Rust]
if (*e).kind == 1 { eval((*e).v.e) }           // before
\end{lstlisting}
\vspace{-1em}
\begin{lstlisting}[language=Rust]
                                               // after
if let C2RustUnnamed::e1(ref x) = (*e).v { eval(*x) }
\end{lstlisting}
\vspace{-0.5em}

\noindent
We handle the disjunction of multiple equality comparisons using \C{|}, the
\emph{or} operator in pattern matching. An example is shown below:

\vspace{-0.5em}
\begin{lstlisting}[language=Rust]
if (*e).kind == 2 || (*e).kind == 3 { eval((*e).v.b.l) }
                                               // before
\end{lstlisting}
\vspace{-1em}
\begin{lstlisting}[language=Rust]
if let C2RustUnnamed::b2(ref x) |              // after
  C2RustUnnamed::b3(ref x) = (*e).v { eval((*x).l) }
\end{lstlisting}
\vspace{-0.5em}

\noindent
However, the idiomatic approach is not applicable to other kinds of conditions,
such as conjunctions and inequalities, because Rust's patterns currently cannot
represent these conditions.
A conjunction specifies that the tag is a particular integer and also that some
boolean formula is satisfied.
When using \C{match} for pattern matching, Rust offers \emph{match
guards}~\cite{rust-match-guard} to express such logic.
However, when using \C{if-let}, this logic is supported through \emph{let
chains}~\cite{rust-let-chain}, which is currently an unstable feature and
requires a special flag to enable.
For this reason, we chose not to transform conjunctions into pattern matching in
this work, leaving it for future work once this feature is stabilized.
On the other hand, an inequality specifies that the tag is not a particular
integer.
However, the purpose of pattern matching is to check whether a value conforms to
a certain pattern, not that it does not, and thus it cannot replace
inequalities.
Thus, the following code is transformed using the na\"ive approach:

\vspace{-0.5em}
\begin{lstlisting}[language=Rust]
if (*e).kind == 1 && ... { eval((*e).v.e) }
if (*e).kind != 1 { ... } else { eval((*e).v.e) }
\end{lstlisting}
\vspace{-0.5em}

The idiomatic transformation is also not applicable when union fields are
accessed without using \C{match} or \C{if} to check the tag field.
C programmers sometimes assume that a tag field has a specific value at a
certain point based on their understanding of the program's behavior and
directly access the union field without checking the tag.
Since tag check is not performed, we cannot transform such code into pattern
matching and must resort to the na\"ive transformation.

The idiomatic transformation consolidates multiple assignment expressions within
a single code block into a single assignment expression that constructs a tagged
union value if the assignments set the tag and union fields.
Below illustrates this transformation:

\vspace{-0.5em}
\begin{lstlisting}[language=Rust]
{ ... (*e).kind = 1; (*e).v.e = ...; ... }     // before
\end{lstlisting}
\vspace{-1em}
\begin{lstlisting}[language=Rust]
{ ... (*e).v = C2RustUnnamed::e1(...); ... }   // after
\end{lstlisting}
\vspace{-0.5em}

\noindent
When multiple assignments are distributed across different code blocks, they are
individually transformed using the na\"ive approach.

\section{Evaluation}
\label{sec:evaluation}

In this section, we evaluate our approach with 36 real-world C programs.
We first describe our implementation of Urcrat, which realizes the proposed
approach (\cref{sec:evaluation:implementation}), and the process of collecting
the benchmark programs (\cref{sec:evaluation:benchmark}).
The implementation and benchmark programs are publicly
available~\cite{hong_2024_13373683}.
We then assess our approach by addressing the following research questions:
\begin{itemize}
  \item RQ1. Precision and recall: Does it identify tag fields without false positives or
    false negatives? (\cref{sec:evaluation:precision})
  \item RQ2. Correctness: Does it transform code while preserving its semantics?
    (\cref{sec:evaluation:correctness})
  \item RQ3. Efficiency: Does it efficiently analyze and transform
    programs? (\cref{sec:evaluation:efficiency})
  \item RQ4. Code characteristics: How much does the code change due to the
    transformation, and how frequently are the helper methods called?
    (\cref{sec:evaluation:code})
  \item RQ5. Impact on performance: What is the effect of replacing unions with
    tagged unions on program performance? (\cref{sec:evaluation:performance})
\end{itemize}
Our experiments were conducted on an Ubuntu machine with Intel Core i7-6700K (4
cores, 8 threads, 4GHz) and 32GB DRAM.
Finally, we discuss potential threats to validity
(\cref{sec:evaluation:threats}).

\subsection{Implementation}
\label{sec:evaluation:implementation}

We built Urcrat on top of the Rust compiler~\cite{rustc}.
Urcrat analyzes Rust code after it has been lowered to Rust's mid-level
intermediate representation (MIR)~\cite{mir}, which expresses functions as
control flow graphs with basic blocks.
For code transformation, Urcrat utilizes Rust's high-level intermediate
representation (HIR)~\cite{hir}, akin to abstract syntax trees but with
syntactic sugar removed and symbols resolved.
We employed C2Rust v0.18.0 with minor modifications.

\subsection{Benchmark Program Collection}
\label{sec:evaluation:benchmark}

\renewcommand{\arraystretch}{0.85}
\begin{table}[t]
  \caption{Benchmark programs}
  \label{tab:benchmark}
  \centering
\footnotesize
  \vspace{-1.5em}
  \begin{tabular}{c|r|r|r|r|r}
    Name&C LOC&Rust LOC&\#Unions&\#Candidates&\#Identified\\\hline
    bc-1.07.1          &10810&16982 &4 &1 &1\\
    binn-3.0$^{**}$    &5686 &4298  &1 &1 &0\\
    brotli-1.0.9$^{**}$&13173&127691&6 &4 &0\\
    cflow-1.7          &20601&26375 &5 &4 &3\\
    compton$^{*}$      &8748 &14084 &2 &2 &2\\
    cpio-2.14          &35934&80929 &10&4 &3\\
    diffutils-3.10     &59377&95835 &7 &5 &4\\
    enscript-1.6.6     &34868&78749 &9 &5 &3\\
    findutils-4.9.0    &80015&139858&13&6 &3\\
    gawk-5.2.2         &58111&140566&17&10&3\\
    glpk-5.0           &71805&145738&18&14&3\\
    gprolog-1.5.0      &52193&74381 &5 &2 &0\\
    grep-3.11          &64084&84902 &11&9 &6\\
    gsl-2.7.1          &227199&422854&14&14&0\\
    gzip-1.12          &20875&21605 &4 &2 &1\\
    hiredis$^{*}$      &7305 &14042 &1 &1 &1\\
    make-4.4.1         &28911&36336 &1 &1 &1\\
    minilisp$^{*}$     &722  &2149  &1 &1 &1\\
    mtools-4.0.43      &18266&37021 &2 &1 &0\\
    nano-7.2           &42999&74994 &6 &4 &3\\
    nettle-3.9         &61835&82742 &5 &2 &1\\
    patch-2.7.6        &28215&103839&3 &1 &1\\
    php-rdkafka$^{*}$  &3771 &28864 &1 &1 &1\\
    pocketlang$^{*}$   &14267&41439 &4 &3 &3\\
    pth-2.0.7          &7590 &12950 &1 &1 &1\\
    raygui$^{*}$       &1588 &17218 &1 &1 &1\\
    rcs-5.10.1         &28286&36267 &1 &1 &1\\
    screen-4.9.0       &39335&72199 &1 &1 &0\\
    sed-4.9            &48190&68465 &8 &7 &4\\
    shairport$^{*}$    &4995 &10118 &2 &1 &1\\
    tar-1.34           &66172&134972&16&12&9\\
    tinyproxy$^{*}$    &5667 &12825 &5 &2 &2\\
    twemproxy$^{*}$    &22738&74593 &8 &7 &5\\
    uucp-1.07          &51123&77872 &3 &3 &0\\
    webdis$^{*}$       &14369&29474 &2 &2 &2\\
    wget-1.21.4        &81188&192742&6 &5 &4\\\hline
\textbf{Total} &&&204&141&74
  \end{tabular}
  \\
  {\scriptsize $^{*}$: from GitHub, $^{**}$: from \textsc{Crown}}
  \vspace{-1.5em}
\end{table}

We collected benchmark programs from three sources:
(1) \textsc{Crown}~\cite{zhang2023ownership}, (2) GNU packages~\cite{gnu}, and
(3) GitHub.
Only 2 out of 20 programs used by \textsc{Crown} have candidate unions, so we
expanded the benchmarks with additional sources.
We chose GNU packages for their representative C projects and GitHub for its
diverse code patterns.
We avoided large codebases because C2Rust often
produces Rust code with type errors, primarily due to missing type casts, which
require significant manual effort to correct.
From GNU, we gathered C packages with less than 250k LOC and individual
Wikipedia entries, indicating they are well-known.
From GitHub, we gathered C projects with less than 1 MB of code and over 1,000
stars.
In both collections, we retained programs (1) compiled successfully on Ubuntu,
(2) transpiled successfully by C2Rust, and (3) containing candidate unions.
This resulted in 24 programs from GNU and 10 from GitHub, giving us a total of
36 benchmark programs.
Of these, 20 required manual fixes after C2Rust's translation, with an average
of 26.6 lines modified.
Columns 2--5 of \cref{tab:benchmark} present the C LOC, Rust LOC, number of
unions, and number of candidate unions in each program, respectively.
In the benchmarks, 21\% of unsafe functions involve unions, and 11\% involve
tagged unions.

\subsection{RQ1: Precision and recall}
\label{sec:evaluation:precision}

We evaluate the precision and recall of the proposed approach.
We first identified the tag field of each candidate union through static
analysis.
Column 6 of \cref{tab:benchmark} shows the number of candidate unions for which
a tag field is identified in each benchmark program.
Out of 141 candidates across 36 programs, Urcrat identifies tag fields for 74
candidates in 29 programs.
We then manually inspected each candidate union to determine the presence of a
tag field, checking for false positives (a tag field identified for a union that
does not have one) and false negatives (no tag field identified for a union that
has one).
For practical use of our tool, false positives are problematic because they
change the program's semantics.
Conversely, false negatives are less problematic as they only prevent the
replacement of unions with tagged unions, and the semantics remains correct.

Our manual inspection shows that the static analysis is precise,
revealing no false positives and only five false negatives.
This results in a precision of 100\% and a recall of 93.7\% ($=74/79$).
Specifically, two false negatives occur in \C{glpk-5.0}, while the others are in
\C{gawk-5.2.2}, \C{screen-4.9.0}, and \C{uucp-1.07}, respectively.
Three of these (from \C{glpk-5.0}, \C{gawk-5.2.2}, and \C{screen-4.9.0}) are due
to intermediate states involving tag values that are not associated with any union fields
by \C{CollectFromAccesses}.
We now discuss the reasons for the remaining two false negatives:

\paragraph{glpk-5.0}

The false negative arises from \C{goto} in the C code:

\vspace{-0.5em}
\begin{lstlisting}[language=C]
switch (tab->type) { case 112: goto input_table;
  case 119: goto output_table; default: abort(); }
input_table: ... return; output_table: ...
\end{lstlisting}
\vspace{-0.5em}

\noindent
In this code, \C{type} is the tag field.
Since Rust does not support \C{goto}, C2Rust translates the code as follows:

\vspace{-0.5em}
\begin{lstlisting}[language=Rust]
match (*tab).type {
 112 => current_block = 1, 119 => { ... } _ => abort(),}
if current_block == 1 { ... }
\end{lstlisting}
\vspace{-0.5em}

\noindent
The resulting code has a variable, \C{current\_block}, which mimics \C{goto}'s
effect.
As the analysis is path-insensitive, it concludes that both \C{112} and \C{119}
are possible values in the true branch of \C{if}, hindering the identification
of the tag field.
To address this issue, we need to either analyze the original C code or modify
the C code to avoid \C{goto}.
We chose the latter and revised the code as follows:

\vspace{-0.5em}
\begin{lstlisting}[language=C]
switch (tab->type) { case 112: ... break;
  case 119: ... break; default: abort(); }
\end{lstlisting}
\vspace{-0.5em}

\noindent
Urcrat can identify the tag field from the modified version.

\paragraph{uucp-1.07}

This false negative is due to a bug in the C code:

\vspace{-0.5em}
\begin{lstlisting}[language=C]
if (qport->uuconf_ttype == 5) {
 if (qport->uuconf_u.uuconf_stli.zdevice != NULL)
  fprintf(e,"%s",qport->uuconf_u.uuconf_smodem.zdevice);
\end{lstlisting}
\vspace{-0.5em}

\noindent
Here, \C{uuconf\_ttype} is the tag field, and the union field associated with
\C{5} is \C{uuconf\_stli}.
However, the code erroneously accesses \C{uuconf\_smodem} in the \C{fprintf}
statement.
The function contains multiple \C{fprintf} statements, suggesting that this bug
was likely introduced through copy-pasting.
After correcting the code to access \C{uuconf\_stli}, Urcrat successfully
identifies the tag field.

\subsection{RQ2: Correctness}
\label{sec:evaluation:correctness}

We evaluate the correctness of our approach by checking whether the transformed
program is compilable and exhibits the same behavior as the original.
For the experiments, we used the fixed code for \C{glpk-5.0} and \C{uucp-1.07},
allowing the identification of additional tag fields.
Consequently, we examined 30 programs:
29 identified without code fixes and 1, which is \C{uucp-1.07}, identified only
after the code fix.
All 30 programs are compilable after transformation.
Among 23 programs with test suites, 17 pass the tests post-transformation.
Those failed are \C{gawk-5.2.2}, \C{grep-3.11}, \C{make-4.4.1}, \C{minilisp},
\C{twemproxy}, and \C{wget-1.21.4}.

We manually investigated the reasons for the failures and found that only
\C{gawk-5.2.2} and \C{twemproxy}'s failures result from imprecise identification
of tag values in the static analysis.
The other failures are due to two specific C code patterns requiring minor
manual code fixes after transformation:
(1) reading a union field other than the last-written one, and
(2) code relying on memory layout.
We now discuss the failure reasons for each program.

\paragraph{gawk-5.2.2}

The failure arises from tag values not being identified due to intraprocedural
analysis.

\vspace{-0.5em}
\begin{lstlisting}[language=C]
INSTRUCTION *bcalloc(int op, ...) {
 INSTRUCTION *cp = ...; cp->opcode = op; ... return cp;}
\end{lstlisting}
\vspace{-0.5em}

\noindent
In this code, \C{opcode} is the tag field, assigned the argument value.
The analysis cannot identify values at the call-site of \C{bcalloc} as possible
\C{opcode} values.
This causes the transformed program to panic when an unknown tag value is passed
to \C{set\_opcode}.

\paragraph{twemproxy}

The failure is due to the lack of C library modeling, preventing identification
of tag values.

\vspace{-0.5em}
\begin{lstlisting}[language=C]
getaddrinfo(n, s, &h, &ai); si->family = ai->ai_family;
\end{lstlisting}
\vspace{-0.5em}

\noindent
Here, \C{family} is the tag field, and \C{ai\_family} determines its value, set
by the \C{libc} function \C{getaddrinfo}.
While \C{2} and \C{10} are possible values, the analysis fails to recognize them
as tag values.
This issue can be resolved by incorporating library modeling into the analysis.

\paragraph{grep-3.11}

It fails because of reading a field not lastly written.

\vspace{-0.5em}
\begin{lstlisting}[language=C]
fetch_token(token, input, syntax);
c = token->opr.c; if (token->type == 2) return;
\end{lstlisting}
\vspace{-0.5em}

\noindent
In this code, \C{type} is the tag field, and \C{opr} is the union value.
The code first reads \C{c} and then checks \C{type}, causing a panic when
\C{get\_c} is called in the transformed code.
Swapping the order of these statements allows the transformed program to pass
the tests.

\paragraph{make-4.4.1}

This is also due to reading a field not lastly written.

\vspace{-0.5em}
\begin{lstlisting}[language=C]
struct function_table_entry { ..., int alloc_fn;
  union {fptr1 func_ptr; fptr2 alloc_func_ptr;} fptr;};
if (!entry_p->fptr.func_ptr) abort();
\end{lstlisting}
\vspace{-0.5em}

\noindent
The union has two fields, \C{func\_ptr} and \C{alloc\_func\_ptr}, with
\C{alloc\_fn} as the tag field.
The code checks if \C{func\_ptr} is null regardless of \C{alloc\_fn}'s value,
exploiting that different function pointer types have the same size and null
representation.
After transformation, this causes a panic if \C{alloc\_func\_ptr} is the
last-written field.
The code can be fixed as follows to pass the tests post-transformation:

\vspace{-0.5em}
\begin{lstlisting}[language=C]
if (entry_p->alloc_fn == 0 && !entry_p->fptr.func_ptr ||
 entry_p->alloc_fn==1 && !entry_p->fptr.alloc_func_ptr)
\end{lstlisting}
\vspace{-0.5em}

\paragraph{minilisp}

This failure is caused by different memory layouts of unions and tagged unions.

\vspace{-0.5em}
\begin{lstlisting}[language=C]
struct Obj { int type; int size; union { ... }; };
Obj *alloc(size_t size, ...) { size += 8; ... }
\end{lstlisting}
\vspace{-0.5em}

\noindent
The \C{alloc} function allocates an \C{Obj} in a global byte array.
It determines the memory size by taking the size of a union field and adding
\C{8} to account for the offset of the union within \C{Obj}.
After transformation, the code becomes as follows:

\vspace{-0.5em}
\begin{lstlisting}[language=C]
struct Obj {size: i32, c2rust_unnamed: C2RustUnnamed_1}
\end{lstlisting}
\vspace{-0.5em}

\noindent
Now, each variant value is at offset \C{16} due to alignment requirements,
resulting in \C{alloc} allocating insufficient memory.
We corrected the code to allocate larger memory, enabling the transformed
program to pass the tests.

\paragraph{wget-1.21.4}

It also stems from different memory layouts.

\vspace{-0.5em}
\begin{lstlisting}[language=C]
addr->family = 2; memcpy(&addr->data, tmp, 4);
\end{lstlisting}
\vspace{-0.5em}

\noindent
Here, \C{family} is the tag field, and \C{2} is associated with the union field
\C{d4} of \C{addr->data}.
The C program writes to \C{addr->data} because \C{addr->data} and
\C{addr->data.d4} denote the same address.
However, after transformation, \C{memcpy} overwrites the tag at offset \C{0} of
\C{addr->data}.
We could pass the tests by fixing the code as follows, facilitating
\C{deref\_d4\_mut} to be called in the transformed code:

\vspace{-0.5em}
\begin{lstlisting}[language=C]
addr->family = 2; memcpy(&addr->data.d4, tmp, 4);
\end{lstlisting}
\vspace{-0.5em}

Currently, for end users of the tool, detecting and fixing incorrect
translations is challenging.
To identify incorrect translations, they must manually inspect the code or run
tests.
If test suites do not exist, creating new test cases can be costly.
Additionally, test cases may not always reveal incorrect translations.
Fixing incorrect code is even more challenging, as users need to identify the
root cause.
The difficulty of this process depends on the root cause.
When tag values are not correctly identified  or fields not lastly written are
read, investigating the cause is relatively straightforward.
The test fails due to a panic, allowing users to check which tag value or field
read triggered the panic.
However, issues related to different memory layouts are more difficult to
diagnose since they do not trigger panics.
In such cases, users must rely on their debugging skills.
We believe it would be beneficial to identify common incorrect translation
patterns and design analyses to detect these patterns, thereby providing users
with warnings.
We leave this as future work.

\subsection{RQ3: Efficiency}
\label{sec:evaluation:efficiency}

\begin{figure}[t]
  \includegraphics[width=0.8\columnwidth]{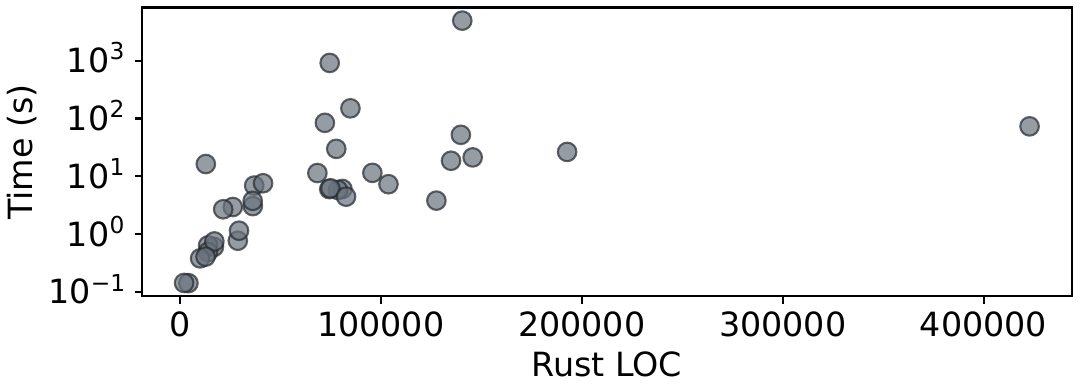}
  \vspace{-1.5em}
  \caption{Urcrat execution time}
  \vspace{-0.5em}
  \label{fig:time}
\end{figure}

We evaluate the efficiency of the proposed approach by measuring the execution
time of Urcrat, which includes both analysis and transformation, for each
program.
\cref{fig:time} presents the execution time relative to the Rust LOC, with the
y-axis displayed in a log scale due to the wide range of execution times.
Urcrat efficiently handles most programs, with 31 programs taking less than a
minute.
The longest execution time is 4,910 seconds for \C{gawk-5.2.2}.
Execution time shows only a weak correlation with code size, as other factors,
primarily the complexity of pointer graphs and the number of analyzed functions,
also significantly influence execution time.

We also investigate the effectiveness of our selective function analysis in
terms of efficiency.
For each program, the transformation consumes less than 1\% of the total time,
as it involves only tree-walking, while the analysis occupies the remaining
time.
This highlights the importance of reducing analysis time.
We calculated the percentage of analyzed functions relative to the total
functions in each program, finding a geometric mean of 4.80\%.
In 28 programs, the analyzed functions constitute less than 10\% of the total
functions.
The highest percentage is 58.23\% in the case of \C{minilisp}.
These results indicate that our approach is effective for efficiency.

\subsection{RQ4: Code Characteristics}
\label{sec:evaluation:code}

To understand the characteristics of the transformed code, we first evaluate the
extent of code changes introduced.
We measured the changes in the 30 programs where tag fields are identified.
On average, the transformation added 861.8 lines per program due to helper
method definitions.
Excluding these, an average of 252.4 lines were inserted and 301.8 lines were
deleted per program.
These results demonstrate the practical utility of our approach, as manually
implementing such significant code modifications would be both time-consuming
and error-prone.

We also evaluate the applicability of the idiomatic transformation, which avoids
inserting method calls.
With only the na\"ive transformation, each program has 194.6 method calls on
average.
In contrast, applying the idiomatic transformation alongside the na\"ive
transformation reduces this to 124.8 method calls per program, achieving a 36\%
decrease.
Specifically, 38.3 calls are removed by replacing \C{match} on tag values with
\C{match} on tagged unions, 15.8 calls by replacing \C{if} with \C{if-let}, and
15.6 calls by consolidating separate assignments into tagged union construction.
The results indicate that the idiomatic transformation improves code quality.

\subsection{RQ5: Impact on Performance}
\label{sec:evaluation:performance}

%

We evaluate the impact of replacing unions with tagged unions on the performance
of the translated programs.
We compare the performance of each Rust program before and after the
transformation, measured in terms of the execution time of the test suite.
%
%
To ensure reliable results, we excluded test suites with execution times shorter
than 0.1 seconds, resulting in 20 programs, and computed the average execution
time from fifty runs for each program.
%

The experimental results show that the performance overhead is negligible.
On average, the transformed programs were only 0.01\% slower than the
original ones.
Among the 20 programs, 5 ran slightly slower after transformation,
while the others were faster.
%
%
We also performed one-sided t-test with the null hypothesis that the program
after transformation is slower than the original program by at least 1\%.
With a significance level of 0.05, we could reject the null hypothesis for 9 out
of 20 programs, implying that the transformation is unlikely to incur observable
performance overhead for them.
For the remaining 11 programs, we could not reject the null hypothesis, but this
does not necessarily mean they have an overhead greater than 1\%.
Repeating the experiments may provide stronger statistical significance,
rejecting the null hypothesis.

\subsection{Threats to Validity}
\label{sec:evaluation:threats}

Threats to external validity primarily stem from the choice of benchmarks.
GNU packages often share common code patterns, which may introduce bias.
Although we included projects from GitHub to enhance diversity, this may not
fully represent the entire C ecosystem.
Notably, large codebases might exhibit different characteristics.
Further experiments with a broader range of C programs would provide greater
confidence in the generalizability of our approach.

The type errors produced by C2Rust also pose a threat to external validity.
Before using our tool, all type errors in C2Rust-generated code must be
resolved.
Larger programs tend to exhibit more type errors after C2Rust's translation,
requiring significant effort to correct.
This could discourage users from adopting both C2Rust and Urcrat, potentially
impeding the widespread adoption of our approach.
To mitigate this issue, we can improve C2Rust or develop techniques for fixing
type errors in C2Rust-generated code.

Threats to construct validity arise from using test suites for correctness and
performance evaluation.
Passing all tests does not guarantee correctness.
However, tests are the most widely used method for practical semantics
validation and successfully discovered incorrect behavior in some programs
during our evaluation.
For performance assessment, the used test suites may be inadequate as they were
not designed for performance measurement.

\section{Related Work}
\label{sec:related}

\paragraph{Improving C2Rust-Generated Code}

Several studies have proposed techniques to improve C2Rust's translation by
replacing unsafe features with safe counterparts in Rust.
However, none of these studies address the translation of unions, which is the
focus of this work.
\textsc{Laertes}~\cite{emre2021translating, emre2023aliasing} and
\textsc{Crown}~\cite{zhang2023ownership} aim to replace raw pointers with
references.
\textsc{Laertes} uses compiler feedback to determine which pointers to replace
and the appropriate lifetimes for the references.
In contrast, \textsc{Crown} employs ownership analysis to facilitate the
replacement of a larger number of pointers compared to \textsc{Laertes}.
Concrat~\cite{hong2023concrat} targets the replacement of some external function
calls with equivalent functions from the Rust standard library.
It employs dataflow analysis to identify the use of locks, allowing for the
substitution of the C lock API with the Rust lock API.

\paragraph{C-to-Rust Translation with Language Models}

Recent studies have developed language models capable of translating
code~\cite{roziere2020unsupervised, lachaux2024dobf, roziere2022leveraging,
szafraniec2023code, liu2023syntax, wang2021codet5, chen2021evaluating,
feng2020codebert, guo2021graphcodebert, touvron2023llama} or proposed techniques
to utilize pre-trained language models for code
translation~\cite{yang2024exploring, pan2024lost}.
Most focus on translating between languages like C++, Java, and Python, rather
than C to Rust.
A notable exception is \citet{roziere2022leveraging}, which translates C++ to
Rust, but their evaluation uses only small competitive programming solutions
that do not involve unions, with only 21\% translated correctly.
Similarly, \citet{pan2024lost} use GPT-4 for C-to-Rust translation, applying it
to CodeNet~\cite{puri2021codenet}, which consists of programming problem
solutions, achieving 61\% correctness.
These results indicate that language models frequently generate code that is
uncompilable or semantically incorrect, while our approach consistently produces
compilable and mostly correct code.

\paragraph{Must-Points-To Analysis}

Researchers have studied must-points-to analysis, but their techniques differ
from ours in purpose, method, and target language.
Our analysis aims to precisely compute struct field values.
In contrast, most studies use must-points-to analysis to enhance the precision
of other analyses.
Altucher and Landi~\cite{altucher1995extended} use must-points-to relations to
compute def-use relations.
Ma et al.~\cite{ma2008computing} and Fink et al.~\cite{fink2008effective}
improve may-points-to relations using must-points-to relations, enabling more
precise detection of null pointer dereferences and Java typestate checking,
respectively.
Nikoli\'c and Spoto~\cite{nikolic2012definite} use must-points-to analysis to
improve other analyses, such as nullness and termination analyses.
While our approach uses may-points-to analysis as a pre-analysis, some
techniques compute may- and must-points-to relations simultaneously.
Emami et al.~\cite{emami1994contexta} conduct interprocedural may- and
must-points-to analysis for compiler optimizations and parallelizations.
Sagiv et al.~\cite{sagiv1999parametric} propose a parametric framework that
generates a family of shape analyses based on three-valued logic, expressing
must-, must-not-, and may-points-to relations.
Unlike most studies, including ours, which focus on imperative languages,
Jagannathan et al.~\cite{jagannathan1998single} propose must-points-to analysis
for functional languages, with results used for optimizations such as closure
conversion.
Techniques applicable to general must-points-to analysis have also been
explored.
Balatsouras et al.~\cite{balatsouras2017datalog} present a declarative model in
Datalog that expresses a wide range of must-points-to analyses.
Kastrinis et al.~\cite{kastrinis2018efficient} propose an efficient data
structure for pointer graphs.

\section{Conclusion}
\label{sec:conclusion}

In this work, we address the challenge of transforming unions with tags into
tagged unions in C2Rust-generated code.
This necessitates identifying tag fields for unions and the tag values
associated with union fields.
To achieve this, we propose a static analysis method that includes
must-points-to analysis for computing struct field values and a heuristic for
interpreting this information.
In addition, we present a code transformation technique that generates idiomatic
code for specific patterns while ensuring semantics preservation for the
remaining code.
Our evaluation shows that the proposed approach is precise, mostly correct, and
scalable.

\begin{acks}

This research was supported by the National Research Foundation of Korea (NRF)
(2022R1A2C200366011 and 2021R1A5A1021944), the Institute for Information \&
Communications Technology Planning \& Evaluation (IITP) grant funded by the
Korea government (MSIT) (2022-0-00460, 2023-2020-0-01819, and 2024-00337703),
and Samsung Electronics Co., Ltd (G01210570).

\end{acks}

\bibliographystyle{ACM-Reference-Format}
\bibliography{references}

\end{document}